\documentclass[twocolumn,showpacs,aps,prb]{revtex4}
\usepackage{epsfig}
\usepackage{graphicx,graphics}
\usepackage{amsmath,ulem}
\usepackage{amssymb}
\usepackage{color}
\usepackage{bm}
\usepackage{subfigure}

\begin{document}

\title{Exact electronic bands for a periodic P\"oschl-Teller potential}

\author{Francesco Di Filippo}
\affiliation{CNR-SPIN, I-84084 Fisciano (Salerno), Italy and
	Dipartimento di Fisica ``E. R. Caianiello'', Universit\`a di
	Salerno, I-84084 Fisciano (Salerno), Italy}

\author{Canio Noce}
\affiliation{CNR-SPIN, I-84084 Fisciano (Salerno), Italy and
	Dipartimento di Fisica ``E. R. Caianiello'', Universit\`a di
	Salerno, I-84084 Fisciano (Salerno), Italy}

\begin{abstract}
We show that supersymmetry is a simple but powerful tool to exactly solve quantum mechanics problems. Here, the supersymmetric approach is used to analyse a quantum system with periodic P\"oschl-Teller potential, and to find out the exact energy spectra and the corresponding band structure.
\end{abstract}


\maketitle

\section{Introduction}

Schr\"odinger equation is certainly one of the cornerstones of theoretical physics.\cite{schroedinger26} This equation is the core equation of quantum mechanics, and it can be considered as the analogue of Newton's second law of classical mechanics. It is the starting point for every quantum mechanical system we want to describe: electrons, atoms, whatever. Thus,  the search for exactly solvable and integrable potentials of this equation has a great significance, because the qualitative understanding of a complicated realistic system can be acquired by analysing exactly solvable simplified models that still retain the essential features of the physical system under investigation. We point out that by exactly solvable model we mean that the eigenvalues and eigenfunctions of the Schr\"odinger equation can be given in an explicit and closed form.\cite{cohen78} 

Several methods and techniques have been developed during the last decades to exactly solve the Schr\"odinger equation or to obtain approximate analytical solutions, with the aim to get better understanding of its dynamical behaviour.\cite{shifman89} Despite all the efforts that have been done, the exact solutions of that equation are limited to a small set of special potentials such as, for instance, the ones considered in the well-known problems of the harmonic oscillator and the hydrogen atom. When periodic potentials are concerned, the situation is even worse. Indeed, the general analysis of the electronic levels in a periodic potential, independently of its form, can be successfully carried out only in one dimension. Although the one dimensional case is in many respects atypical or irrelevant, some of the features of the three dimensional band structure may be described through approximate calculations emerging from an exact treatment in one dimension. Also in this case, one is able to carry out an exact calculation of the allowed and forbidden bands for very special models as the Dirac comb or the Kronig-Penney model.\cite{griffiths05} Thus, it would be valuable to investigate more general and realistic potentials even though one usually does this job by approximate methods based on elementary perturbation theory. Routinely, one considers the weak and tight binding approximations starting out, in the first case, with the wave function for a free electron, and in the other one with the wave function of an electron bound to an atom. 

Recently, the concept of supersymmetry (SUSY) has been profitably applied to many non-relativistic quantum mechanical problems, paving a route to enlarge the class of exactly solvable Schr\"odinger equations.\cite{cooper95,junker96,bagchi01,sukumar04,cooper05,david10} In particular, there is now a much deeper understanding of why Schr\"odinger equation with certain potentials may be analytically solved and several powerful new approximate methods have been formulated for handling potentials which are not exactly solvable. In particular, the exact solution of the Schr\"odinger equation for solvable potentials can be understood in terms of few basic ideas which include supersymmetric partner potentials, shape invariance and operator transformations.\cite{cooper95} For completeness, we notice that standard textbook solvable potentials all exhibit the property of shape invariance.\cite{schwabl}
In the case of periodic potentials, the supersymmetric properties slightly change compared to then strategy adopted for ordinary potentials,\cite{dunne98,khare04,ioffe08} the main difference being related to the finiteness of the domain of the potential. Indeed, in this case the potentials indefinitely repeated, so that one can try to solve the Schr\"odinger equation considering potentials whose shape looks more realistic than the Dirac comb or the Kronig-Penney potential.
The aim of this paper is to present an exact solution of the Schr\"odinger equation for a periodic P\"oschl-Teller potential, showing that the  SUSY concept turns out an useful and powerful tool helping us to find out the solution of this equation. 

The paper is organized as follows: we start by introducing, in Sec. II, the general properties of the SUSY within the quantum mechanics realm, pointing out important concepts such as Hamiltonian hierarchy and shape invariance. In Sec. III, we present and discuss the exact solution of the Schr\"odinger equation for a periodic P\"oschl-Teller potential. Sec. IV is devoted to final remarks and conclusions.

\section{General properties of SUSY}
Let us consider a time independent, one dimensional, quantum system described by an Hamiltonian $H_{1}$.\cite{gendenshtein83} Without loss of generality, we can suppose that the ground state of the system has zero energy. 

\noindent The Schr\"odinger equation for the ground state is
\begin{equation}
H_{1}\psi_{0}=-\frac{\hbar^{2}}{2m}\frac{d}{dx}\psi_{0}+V_{1}\left(x\right)\psi_{0}=0\, .
\end{equation}

\noindent Then, let us suppose that there exist two operators $A$ and $A^{\dagger}$ that factorize
$H_{1}$, i.e.
\begin{equation}
H_{1}=A^{\dagger}A\, .
\end{equation}
In this case, the following operators  
\begin{equation}
A=\frac{\hbar}{2m}\frac{d}{dx}+W\left(x\right)\, ,
\end{equation}
\begin{equation}
A^{\dagger}=-\frac{\hbar}{2m}\frac{d}{dx}+W\left(x\right)\, ,
\end{equation}
reproduce $H_{1}$, provided that $W\left(x\right)$ is a real function such that
\begin{equation}
W^{2}\left(x\right)-\frac{\hbar}{2m}W'\left(x\right)=V_{1}\left(x\right)\, .
\end{equation}
We notice that, changing the order of $A$ and $A^{\dagger}$, we get a new Hamiltonian
operator $H_{2}$
\begin{equation}
H_{2}=AA^{\dagger}\, ;
\end{equation}
$H_{1}$ and $H_{2}$ are called supersymmetric partners.

\noindent Now, let us discuss the properties of these Hamiltonians. To this end, 
in the following we will denote by $\psi_{n}^{\left(i\right)}$ $\left(E_{n}^{\left(i\right)}\right)$
the $n$-th eigenvector (eigenvalue) of the the  Hamiltonian $H_{i}$. \\
\noindent First of all, it can be trivially shown that both Hamiltonians $H_{1}$ and $H_{2}$ admit a set of non-negative eigenvalues. Indeed, 
\begin{equation}
E_{n}^{\left(1\right)}=\left\langle \psi_{n}^{\left(1\right)}\right|A^{\dagger}A\left|\psi_{n}^{\left(1\right)}\right\rangle =\left|A\psi_{n}^{\left(1\right)}\right|^{2}\ge0\label{eq:7}\, ,
\end{equation}
and
\begin{equation}
E_{n}^{\left(2\right)}=\left\langle \psi_{n}^{\left(2\right)}\right|AA^{\dagger}\left|\psi_{n}^{\left(2\right)}\right\rangle =\left|A\psi_{n}^{\left(2\right)}\right|^{2}\ge0\label{eq:8}\, .
\end{equation}

\noindent Moreover, at most one of the supersymmetric partners exhibits a normalized zero energy ground state. This conclusion stems from
Eqs.(\ref{eq:7}-\ref{eq:8}) and the definition of the operator $A$ implying that
\begin{equation}
\psi_{0}^{(1)}(x)=N_1 exp\left\{ -\frac{\sqrt{2m}}{\hbar}\int_{0}^{x}W(\tilde{x})d\tilde{x}\right\} \label{eq:12}\, ,
\end{equation}

\begin{equation}
\psi_{0}^{(2)}(x)=N_2 exp\left\{ \frac{\sqrt{2m}}{\hbar}\int_{0}^{x}W(\tilde{x})d\tilde{x}\right\} \label{eq:13}\, .
\end{equation}
Thus, at least one of the two lowest energy eigenstates is not normalizable.

\noindent Finally, the two Hamiltonians have almost identical energy spectrum, apart from possibly the zero energy level. Indeed, applying $A$ $\left(A^{\dagger}\right)$ to an eigenvector of $H_{1}\,\left(H_{2}\right)$ one gets an eigenvector
of $H_{2}\,\left(H_{1}\right)$, associated with the the same eigenvalue: 
\begin{equation}
AA^{\dagger}A\psi_{n}^{(1)}=AE_{n}^{(1)}\psi_{n}^{(1)}\Rightarrow H_{2}\left(A\psi_{n}^{(1)}\right)=E_{n}^{(1)}\left(A\psi_{n}^{(1)}\right)
\end{equation}
Obviously, this is true unless $A\psi_{n}^{\left(1\right)}$ is zero. 

\noindent Therefore, two scenarios are possible: broken and unbroken SUSY: when the two energy spectra are identical
SUSY is broken, otherwise they differ for the ground state energy and SUSY is unbroken.

In the case of unbroken SUSY, we can go further factorizing $H_{2}$ by means of the operator $A_{2}$ and its adjoint such that
\begin{equation}
H_{2}=A_{2}^{\dagger}A_{2}+E_{0}^{\left(2\right)}\, .
\end{equation}
In this way, we may get a new Hamiltonian, as previously done to define $H_{2}$:
\begin{equation}
H_{3}=A_{2}A_{2}^{\dagger}+E_{0}^{\left(2\right)}\, .
\end{equation}
Iterating this procedure we may finally write down $H_{n}$ as follows
\begin{equation}
H_{n}=A_{n}^{\dagger}A_{n}+\sum_{k=1}^{n}E_{0}^{(k)}=A_{n-1}A_{n-1}^{\dagger}+\sum_{k=1}^{n-1}E_{0}^{(k)}\, .
\end{equation}
Summarizing, starting from an Hamiltonian having $n$ bound states, we can introduce
a chain of $n$ Hamiltonians such that $H_{i}$ and $H_{i+1}$ have an energy spectrum almost identical, apart from the absence of the ground state of $H_{i+1}$ within the energy eigenvalues of $H_{i}$. This result is represented in Fig.\,1.

\begin{figure}[b]
\includegraphics[scale=0.55]{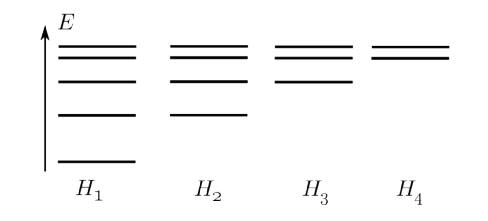}
\caption{Example of a hierarchy of Hamiltonians when four Hamiltonians are considered. The orizontal bars represent the energy levels of the corresponding Hamiltonians $H_i$.}
\end{figure}

From the knowledge of the ground state of $H_{n}$, that can be found using Eq. (\ref{eq:12}), we can find the $n$-th excited state of $H_{1}$ as follows 
\begin{equation}
\psi_{n}^{\left(1\right)}=\frac{1}{\sqrt{E_{1}^{\left(1\right)}...E_{n-1}^{\left(1\right)}}}A_{1}^{\dagger}A_{2}^{\dagger}...A_{n-1}^{\dagger}\psi_{0}^{(n)}\label{hamhie}\, .
\end{equation}

\noindent The Hamiltonian hierarchy turns out very useful when the potential term of the Hamiltonian $V_1$ satisfies 
the following shape invariance condition: 
\begin{equation}
V_{2}\left(x,a_{1}\right)=V_1\left(x,a_{2}\right)+g\left(a_{1}\right)\, .
\end{equation}
Here $a_{1}$ and $a_{2}$ are parameters or set of parameters related between them by some link $a_{1}=f\left(a_{2}\right)$.
In this case the quantum mechanical problem can be easily solved. Indeed, using the following equation

\begin{equation}
H_{2}=-\frac{\hbar^{2}}{2m}\frac{d^{2}}{dx^{2}}+V_{2}(x,a_{1})=-\frac{\hbar^{2}}{2m}\frac{d^{2}}{dx^{2}}+V_{1}(x,a_{2})+g(a_{1})\, ,
\end{equation}
one may iterate the procedure, getting the $n$-th Hamiltonian:

\begin{equation}
H_{n}=-\frac{\hbar^{2}}{2m}\frac{d^{2}}{dx^{2}}+V_{1}(x,a_{n})+\sum_{k=1}^{n-1}g(a_{k})\, .
\end{equation}
We can show that the ground state of $H_{n}$ is $\psi_{0}^{(1)}(x,a_{n})$, and its energy is $\sum_{k=1}^{n-1}g(a_{k})$. Indeed,  

\begin{equation}
\begin{alignedat}
{1}H_{n}\psi_{0}^{(1)}(x,a_{n}) & =\underbrace{\left(-\frac{\hbar^{2}}{2m}\frac{d^{2}}{dx^{2}}+V_{1}(x,a_{n})
	\right)\psi_{0}^{\left(1\right)}(x,a_{n})}_{H_{1}(x,a_{n})\psi_{0}^{1}(x,a_{n})=0}\\
&+\left(\sum_{k=1}^{n-1}g(a_{k})\right)\psi_{0}^{\left(1\right)}(x,a_{n})\\
& =\left(\sum_{k=1}^{n-1}g(a_{k})\right)\psi_{0}^{\left(1\right)}(x,a_{n})\, ,
\end{alignedat}
\end{equation}
trivially proving our claim.

\noindent We want to stress that the ground state energy of $H_{n}$ is equal to the
$n$-th excited level of $H_{1}$, implying that the energy spectrum of $H_1$ is given as follows:

\begin{equation}
\begin{array}{cc}
E_{0}^{(1)}=0\qquad & \qquad E_{n}^{(1)}=E_{0}^{(n+1)}=\sum\limits_{k=1}^{n}g(a_{k})\end{array}\label{eq:spettro Sip}\, .
\end{equation}
Furthermore, we can use Eq.(\ref{hamhie}) to find out the corresponding eigenstates in the following way:

\begin{equation}
\begin{alignedat}
{1} \psi_{n}^{(1)}(x,a_{1})=&\frac{1}{\sqrt{E_{1}^{(1)}\cdot\cdot\cdot E_{n}^{(1)}}} \\ & A^{\dagger}(x,a_{1})A^{\dagger}(x,a_{2})\cdot\cdot\cdot A^{\dagger}(x,a_{n})\psi_{0}^{(1)}(x,a_{n+1})\, .\label{shapein}
\end{alignedat}
\end{equation}

\section{Exact electronic bands for periodic P\"oschl-Teller potential}

\subsection{P\"oschl-Teller potential: non periodic case}

In this Section we will get the energy spectrum for periodic P\"oschl and Teller potential. To this end, we will firstly show how the Schr\"odinger equation may be exactly solved for non-periodic P\"oschl and Teller potential, and then, taking advantage from this solution, we will present the solution for the periodic case. 

\noindent The P\"oschl and Teller potential is given by
\begin{equation}
V=-\frac{\alpha^{2}\hbar^{2}}{2m}\frac{l\left(l+1\right)}{cosh^{2}\,\alpha x}\, ,
\end{equation}
with $l\in N$ and $\alpha \in R$.
We point out that this potential has been introduced by P\"oschl and Teller to study the vibrational spectra of polyatomic molecules.\cite{poschl33}
\noindent For completeness, we would like to stress that the Schr\"odinger equation with the potential of Eq.(22) can be exactly solved also for $l\in R$.\cite{flugge71}

\noindent Although the SUSY allows for an exact solution of the Schr\"odinger equation only for the special case of integer values of $l$, it has the advantage that the solution of this equation may be easily derived and worked with. 
\\

To find out an explicit solution, we start considering a function $W(x)$ of the form:

\begin{equation}
	W\left(x\right)=l\frac{\hbar}{\sqrt{2m}}\alpha\tanh\,\alpha x\, ,
\end{equation}

\noindent with the supersymmetric partners $V_1$ and $V_2$ given by:

\begin{equation}
	V_{1}=\frac{\alpha^{2}\hbar^{2}}{2m}\left(l^{2}-\frac{l(l+1)}{cosh^{2}\alpha x}\right)\, ,
\end{equation}

\begin{equation}
	V_{2}=\frac{\alpha^{2}\hbar^{2}}{2m}\left(l^{2}-\frac{(l-1)l}{cosh^{2}\alpha x}\right)\, .
\end{equation}

\noindent We notice that the shape invariance condition Eq.(16) is satisfied when $a_{1}=l$, $a_{2}=l-1$
and $g(l)=\frac{\alpha^{2}\hbar^{2}}{2m}(2l-1)$.\\
Moreover, since the potential $V_{l+1}$ is a constant, the energy spectrum of $V_{1}$
is made of $l$ bound states, followed by a continuum spectrum for energy values
greater than $\frac{\alpha^{2}\hbar^{2}}{2m}l^{2}$.\\

\noindent According to Eq.\eqref{eq:spettro Sip} the values of the energy of the bound states of $V_{1}$
are obtained as

\begin{equation}
	E_{l,n}=\sum_{\tilde{l}=l}^{l-(n-1)}g(\tilde{l})=\frac{\alpha^{2}\hbar^{2}}{2m}\left(2nl-n^{2}\right)\qquad {\mathrm{for}}\,\,0\leq n<l \, , 
\end{equation}
so that the energy spectrum of $V$ may be easily obtained shifting this spectrum by
$-\frac{\alpha^{2}\hbar^{2}}{2m}l^{2}$. Thus, we get
\begin{equation}
\!\!\!\!\!\!	E_{n}=\frac{\alpha^{2}\hbar^{2}}{2m}\left(2nl-n^{2}-l^{2}\right)=-\frac{\alpha^{2}\hbar^{2}}{2m}(n-l)^{2}\quad 
	{\mathrm{for}}\,\,0\leq n<l \, ,
\end{equation}
while the continuum part of the spectrum is simply given by non negative energies $E>0$.\\

\subsection{P\"oschl-Teller potential: periodic case}

Taking advantage of the results of the previous Section, we now study the periodic case considering the potential given in Eq.(22) for the special case $l=1$, as plotted in Fig.2 for $a=2$ and $\alpha=1$. Looking at the plot, we may observe that, when the motion of a particle in a one-dimensional lattice is concerned, its shape appears more realistic than, for instance, the Dirac comb or the Kronig-Penney potential.

\begin{figure}
	\includegraphics[scale=0.6]{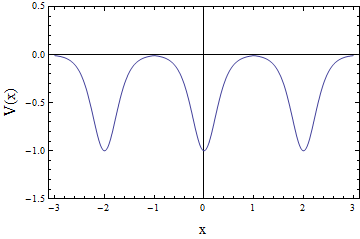} 
	\caption{Shape of one-dimensional periodic P\"oschl-Teller potential. $V(x)$ is measured in units of $\frac{\hbar^{2}}{m}=1$, the lattice constant is $a=2$, and we have assumed $l=1$ and $\alpha=1$.}
\end{figure}

In this case $V$ is given by
\begin{equation}
	V=-\frac{\alpha^{2}\hbar^{2}}{m}\frac{1}{cosh^{2}\,\alpha x}\, .
\end{equation}

\noindent To exactly solve the Schr\"odinger equation for this potential, we first confine the calculation to the interval $\left[-a;a\right]$, and then we apply the Bloch theorem.

\noindent As pointed out before, the function $W(x)$ has the form
\begin{equation}
	W\left(x\right)=\frac{\hbar}{\sqrt{2m}}\alpha\tanh\,\alpha x\, ,
\end{equation}

\noindent so that we have:
\begin{equation}
V_{1}=\frac{\alpha^{2}\hbar^{2}}{2m}\left(1-\frac{2}{cosh^{2}\alpha x}\right) \, ,
\end{equation}
\noindent and 
\begin{equation}
V_{2}=\frac{\alpha^{2}\hbar^{2}}{2m}\, .
\end{equation}

\noindent Since the eigenfunction of the model Hamiltonian with $V=V_2$ are the free particle eigenfunctions, as previously stated, the eigenfunctions of the model Hamiltonian with $V$ given in Eq. (28) can be obtained applying to the free particle eigenfunctions the following operator:
\begin{equation}
A^{\dagger}=\frac{\hbar}{\sqrt{2m}}\left(-\frac{d}{dx}+\alpha\tanh(\alpha x)\right)\, .
\end{equation}
When $E>0$ we get

\begin{equation}
\begin{alignedat}
{1}\psi \left(x,k\right)\propto &-\frac{d}{dx}\phi(x)+\\ &
+\alpha\tanh(\alpha x)\, \phi(x)\, ,
\end{alignedat}
\end{equation}
where 
\begin{equation}
k=\sqrt{\frac{2m}{\hbar^2}E}\, ,
\end{equation}
and
\begin{equation}
\phi(x)=b\cos(kx)+c\sin(kx)\, 
\end{equation}
\noindent are the free particle eigenfunctions.

\noindent For $E<0$ the eigenfunctions may be easily obtained replacing the trigonometric functions with the corresponding hyperbolic ones:

\begin{equation}
\begin{alignedat}
{1} \psi\left(x,k\right)\propto &-\frac{d}{dx}\left[b\cosh(kx)+c\sinh(kx)\right]+\\ &
+\alpha\tanh(\alpha x)\, \left[b\cosh(kx)+c\sinh(kx)\right]\, ,
\end{alignedat}
\end{equation}

\noindent where
\begin{equation}
k=\sqrt{-\frac{2m}{\hbar^2}E}\, .
\end{equation}

\noindent We want to stress that the normalization constant of the eigenfunctions is not relevant since the constraints on $k$, imposed by the continuity of $\psi$ and its derivative, are independent of the normalization constant.

\noindent Applying the Bloch theorem:
\begin{equation}
\psi\left(a\right)=e^{2ia\gamma}\psi\left(-a\right)
\end{equation} 
and considering that
\begin{equation}
\frac{d\psi}{dx}\bigg|_{a}=e^{2ia\gamma}\frac{d\psi}{dx}\bigg|_{-a}
\end{equation} 
we may straightforwardly determine the energy bands. The calculation is quite tedious; details on this procedure are reported in the Appendix.

\noindent To get the energy bands we have to solve the transcendental equation
\begin{equation}
\cos(\gamma a)=f\left(k,\alpha\right)\label{tran}\, ,
\end{equation}

 
\begin{figure}
	\includegraphics[scale=0.6]{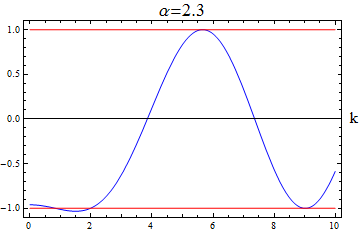} 
	\caption{Band structure for $l=1$, $E>0$, and for $\alpha=2.3$}\label{band1}
\end{figure}

\noindent Let us first discuss the case $E>0$ and $l=1$.
\\ 
\noindent In Fig.(\ref{band1}) we have plotted the function $f\left(k,\alpha\right)$, given in Eq. (A6), choosing $\alpha=2.3$. For this particular choice of $\alpha$,
we observe some energy gaps in the energy region where the function $f\left(k,\alpha\right)$  is either greater than 1 or
smaller than $-1$. Moreover, apart from the first gap, the other forbidden regions are
very small and they may be seen only by zooming in the figure.
For completeness,  we notice that the shape of the function does not
change significantly varying $\alpha$, i. e. only the first gap region may increase or decrease, eventually disappearing, when $\alpha$ is changed. Besides, we do not observe qualitative variations when $l$ increases. In conclusion, even though the shape of this potential may appear more realistic than, for instance, the Dirac comb, for the latter the band structure is more rich and the energy band and the corresponding gaps look more realistic.\cite{griffiths05}

\begin{figure}
	\includegraphics[scale=0.60]{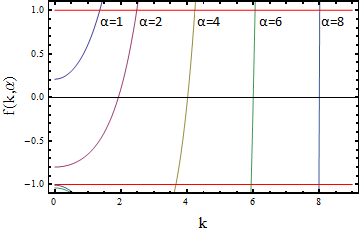} 
	\caption{Band structure for $l=1$, $E<0$, and $\alpha$=1, 2, 4, 6 and 8, from left to right.}\label{band2}
\end{figure}

Let us now study the case $E<0$.
\\
\noindent In Fig.(\ref{band2}) we plot the function $f\left(k,\alpha\right)$ for $l=1$ and  for different values of the parameter $\alpha$, namely  $\alpha$=1, 2, 4, 6 and 8.

\begin{figure}
	\includegraphics[width=7cm]{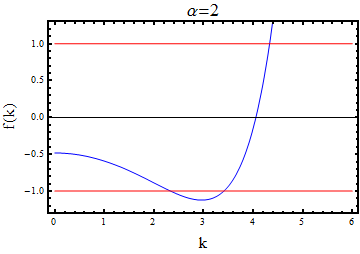}
\end{figure}
\begin{figure}
	\includegraphics[width= 7cm]{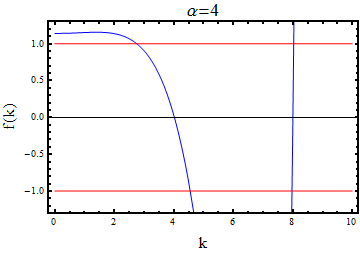}
	\caption{Band structure for $E<0$, $l=2$, and $\alpha=2$ (top panel) and $\alpha=4$ (bottom panel).}\label{band3}
\end{figure}

\noindent In this case, as stressed in the previous section for the non-periodic P\"oschl-Teller potential, the model Hamiltonian exhibits only one bound state corresponding to the energy eigenvalue $E=\frac{\hbar^2}{2m}\alpha^2$. When $\alpha$ increases, all the P\"oschl-Teller wells are separated from each other and the energy level of the periodic case is almost equal to the non periodic case leading to $k=\alpha$. On the other hand, for small values of $\alpha$, the potential wells overlap and the energy levels split into a band. In this case and for $l>1$, the solution of the Schr\"odinger equation would support more than one band. For instance, for larger values of $l$, namely $l=2$, we find two different bands as shown in Fig.(\ref{band3}). Also in this case, the bands are centred at the value corresponding to the energy level of the non periodic case and, for larger values of $\alpha$, the bands shrink into  single levels.

\section{Conclusions}
In this paper we have applied a method which, making use of SUSY combined with Bloch theorem, allows to solve the Schr\"odinger equation for a realistic periodic potential. We have applied this procdure to the periodic P\"oschl-Teller well, explicitly calculating the energy spectrum and the band structure. We have presented the results for both positive and negative energies and we have discussed the results obtained. We point out that the method we have presented turns out to be very efficient and flexible so that it may be applied to other realistic periodic potentials.

We would also like to stress that there are other ways to apply SUSY to the solution of the Schr\"odinger equation in the presence of periodic potentials.  Indeed, it is possible to start from a problem whose band structure is known and find out another potential having the same energy spectrum. For instance, it may be easily shown that the repulsive periodic Dirac comb with P\"oschl-Teller wells between the Dirac deltas has the same band structure of an attractive periodic Dirac comb potential.\cite{oeftiger10} This not trivial conclusion may be inferred simply finding the superpotential that connects the two potentials.

 \renewcommand{\theequation}{A\arabic{equation}}
 \setcounter{equation}{0}
 \section*{APPENDIX}

\noindent Let us first consider the case $E>0$ and $l=1$.

\noindent As discussed in the text,  we start from the eigenfuntion $\psi\left(x\right)$ as reported in Eq.(33): 
{\footnotesize \begin{equation}
\psi\left(x\right)=b\sin(kx)-c\cos(kx)+\frac{\alpha\tanh(\alpha x)(b\cos(kx)+c\sin(kx))}{k}
\end{equation}}
\!for $-a<x<a$.\\

\noindent Then, applying the continuity conditions Eqs.(38)-(39) at $x=a$ for $\psi$; we get: 
\begin{footnotesize}
	\begin{equation}
\begin{alignedat}
{1}&\frac{\alpha\tanh(\alpha a)(b\cos(ak)+c\sin(ak))}{k}+b\sin(ak)-c\cos(ak)= \\ 
& e^{2ia\gamma}\left(-\frac{\alpha tanh(\alpha a)(b\cos(ak)-c\sin(ak))}{k}-b\sin(ak)-c\cos(ak)\right)\, ,
\end{alignedat}
\end{equation}
\end{footnotesize}
\noindent and
{\scriptsize 
\begin{equation} 
\begin{alignedat}
{2}&\!\!\!\!\frac{\alpha^{2}\text{sech}^{2}(a\alpha)(b\cos(ak)+c\sin(ak))}{k}+\frac{\alpha\tanh(a\alpha)(ck\cos(ak)-bk\sin(ak))}{k}+\\
 &\!\!\!\!+bk\cos(ak)+ck\sin(ak)=e^{2ia\gamma}\left[\frac{\alpha^{2}\text{sech}^{2}(a\alpha)(b\cos(ak)-c\sin(ak))}{k}\right] -\\&\!\!\!\!-e^{2ia\gamma}\left[ \frac{\alpha\tanh(a\alpha)(bk\sin(ak)+ck\cos(ak))}{k}+bk\cos(ak)-ck\sin(ak)\right]\, .
\end{alignedat}
\end{equation}}

\noindent Solving these two equations for $\frac{b}{c}$ we get:
{\footnotesize \begin{equation}
\!\!\begin{cases}
\frac{b}{c}= & -\frac{\left(-1+e^{2ia\gamma}\right)(k\cos(ak)-\alpha\tanh(a\alpha)\sin(ak))}{\left(1+e^{2ia\gamma}\right)(\alpha\tanh(a\alpha)\cos(ak)+k\sin(ak))}\\
\\
\frac{b}{c}= & \frac{\left(1+e^{2ia\gamma}\right)\left(k^{2}\sin(ak)+\alpha^{2}\text{sech}^{2}(a\alpha)\sin(ak)+\alpha k\tanh(a\alpha)\cos(ak)\right)}{\left(-1+e^{2ia\gamma}\right)\left(k^{2}\cos(ak)+\alpha^{2}\text{sech}^{2}(a\alpha)\cos(ak)-\alpha k\tanh(a\alpha)\sin(ak)\right)}\, .
\end{cases}
\end{equation}}
\noindent Assuming that $k$ is measured in units of  $\frac{1}{a}$, equating and simplifying the RHS we obtain:
{\scriptsize \begin{equation}
\begin{alignedat}
{1}&k\text{sech}^{2}\left(\frac{\alpha}{2}\right)\Bigg\{\cos(\gamma)\left(\alpha^{2}+k^{2}\right)+\cosh(\alpha)\bigg[\cos(\gamma)\left(\alpha^{2}+k^{2}\right)+\\
&+\left(\alpha^{2}-k^{2}\right)\cos(k)+2\alpha k\tanh\left(\frac{\alpha}{2}\right)\sin(k)\bigg]+\left(-3\alpha^{2}-k^{2}\right)\cos(k)\Bigg\}+\\
&+16\alpha\sinh^{4}\left(\frac{\alpha}{2}\right)\text{csch}^{3}(\alpha)\left(\alpha^{2}+k^{2}\right)\sin(k)=0\, .
\end{alignedat}
\end{equation}}
\noindent Solving this equation for $\cos\,\gamma$, we finally get
{\footnotesize \begin{equation}
\begin{alignedat}
{4} \cos\,\gamma&=\Bigg[k^{3}\cosh(\alpha)\cos(k)+k^{3}\cos(k)-2\alpha k^{2}\cosh(\alpha)\tanh\left(\frac{\alpha}{2}\right)\\
&\sin(k)-16\alpha k^{2}\sinh^{4}\left(\frac{\alpha}{2}\right)\cosh^{2}\left(\frac{\alpha}{2}\right)\text{csch}^{3}(\alpha)\sin(k)-\\
&-16\alpha^{3}\sinh^{4}\left(\frac{\alpha}{2}\right)\cosh^{2}\left(\frac{\alpha}{2}\right)\text{csch}^{3}(\alpha)\sin(k)+3\alpha^{2}k\cos(k)-\\
&-\alpha^{2}k\cosh(\alpha)\cos(k)\Bigg]\frac{1}{k(\cosh(\alpha)+1)\left(\alpha^{2}+k^{2}\right)} \label{f(ka)}\, .
\end{alignedat}
\end{equation}}
\noindent This function has been plotted in Fig. (3).

\noindent The case $E<0$ case is analogous, the only difference being either the substitution of the trigonometric functions with the corresponding  hyperbolic ones or the assumption of imaginary $k$.

\noindent For different $l$, we have to apply the operator $A^{\dagger}$ $l$ times, as specified in \eqref{shapein}, to the free particle eigenfunction Eq.(35). So, for instance, for $l=2$ we get:

{\footnotesize \begin{equation}
\begin{alignedat}
{1}\psi\left( x\right) &= k^2 \big[B \cos (k x)+C \sin (k x)\big] + 2 \alpha ^2 \tanh ^2(\alpha  x)\\& \big[B \cos (k x)+C \sin (k x)\big]-\alpha ^2 \text{sech}^2(\alpha  x) (B \cos (k x)+\\
&+C \sin (k x))-3 \alpha  k \tanh (\alpha  x) \big[C \cos (k x)-B \sin (k x)\big]\, .
\end{alignedat}
\end{equation}}
\noindent Starting from this eigenfunction and following the procedure outlined above, we obtain:

{\footnotesize \begin{equation}
	\begin{alignedat}
{5}f(k,\alpha)&\equiv\Bigg\{k \left(2 \alpha ^2+k^2+3 \alpha  k\right)^2 \cosh \left(k-\frac{5 \alpha }{2}\right)+\\
\!\!\!\!\!\!\!\!\!\!\!\!&+(5 k-12 \alpha ) \left(2 \alpha ^2+k^2+3 \alpha 
	k\right)^2 \cosh \left(k-\frac{3 \alpha }{2}\right)+\\
\!\!\!\!\!\!\!\!\!\!\!\!&+(k-2 \alpha ) \bigg\{2 \left(-15 \alpha ^2+5 k^2-4 \alpha  k\right) (2 \alpha +k)^2 \cosh \left(k-\frac{\alpha }{2}\right)+\\
\!\!\!\!\!\!\!\!\!\!\!\!&+(k-2 \alpha ) \bigg[2
	\left(-30 \alpha ^3+5 k^3+14 \alpha  k^2-7 \alpha ^2 k\right) \cosh \left(\frac{\alpha }{2}+k\right)+\\
\!\!\!\!\!\!\!\!\!\!\!\!&+(k-\alpha )^2 \left((12 \alpha +5 k) \cosh \left(\frac{3 \alpha }{2}+k\right)+k \cosh
	\left(\frac{5 \alpha }{2}+k\right)\right) \bigg]\,\bigg\}\Bigg\}\\
	\!\!\!\!\!\!\!\!\!\!\!\!&\frac{\text{sech}^5\left(\frac{\alpha }{2}\right) }{32 \left(k^5-5 \alpha ^2 k^3+4 \alpha ^4 k\right)}
	\end{alignedat}\label{f(ka)2}\, .
	\end{equation}}
\noindent This function is plotted in Fig.(5).


\end{document}